\documentclass[12pt,a4paper]{article}
\usepackage{amsmath,amsfonts,amssymb,mathrsfs}

\usepackage[english]{babel}

\usepackage{booktabs}
\usepackage{varwidth}
\usepackage{ulem,xcolor}

\usepackage{setspace}

\usepackage{bm,latexsym,euscript,textcomp}
\usepackage{graphicx,color,graphics,epsf,dcolumn}
\usepackage{natbib}
\usepackage{epstopdf}

\usepackage{tikz}
\usetikzlibrary{patterns}
\usetikzlibrary{decorations.text}
\usetikzlibrary{shapes.multipart}
\usetikzlibrary{arrows,shapes.geometric,positioning}
\usetikzlibrary{shapes,positioning,decorations.pathmorphing}
\usetikzlibrary{calc}
\usepackage{booktabs}

\usetikzlibrary{backgrounds}
\usepackage{varwidth}
\usetikzlibrary{fit}

\newcommand{\beq}{\begin{equation}}
\newcommand{\eeq}{\end{equation}}
\newcommand{\pt}{\partial}

\title{Reply to Jaramillo et al. 2019's comments in the Journal of the Atmospheric Sciences \newline doi: 10.1175/JAS-D-19-0025.1}

\author{A. M. Makarieva$^{1,2}$\thanks{\textit{Corresponding author.} {E-mail: ammakarieva@gmail.com}}, 
A. D. Nobre$^3$, \\A.V. Nefiodov$^1$, D.  Sheil$^{4}$,  P. Nobre$^5$, B.-L. Li$^2$}

\begin{document}
\large
\maketitle

\noindent
$^1$Theoretical Physics Division, Petersburg Nuclear Physics Institute, 188300 Gatchina, St. Petersburg, Russia
$^2$USDA-China MOST Joint Research Center for AgroEcology and Sustainability, University of California, Riverside 92521-0124, USA
$^3$Centro de Ci\^{e}ncia do Sistema Terrestre INPE, S\~{a}o Jos\'{e} dos Campos, S\~{a}o Paulo  12227-010, Brazil
$^4$Faculty of Environmental Sciences and Natural Resource Management, Norwegian University of Life Sciences, \AA s, Norway
$^5$Center for Weather Forecast and Climate Studies INPE, S\~{a}o Jos\'{e} dos Campos, S\~{a}o Paulo 12227-010, Brazil

\vspace{0.3cm}
\citet{JMR1} (hereafter JMR1) asserted that our theory, henceforth known as CIAD (condensation-induced atmospheric dynamics), modifies the equation of vertical motion in a manner that violates Newton's third law.
\citet{JMR2} (hereafter JMR2) have subsequently conceded that this is not the case.
This would have resolved the original misunderstanding, had it not been for
new claims in JMR2 that necessitate further correction.

\vspace{0.3cm}
{\bf 1. A "potential" modification.} 
JMR2 state that while CIAD does not modify the equation of vertical motion (indeed 
they were unable to locate any such modified equation), they nevertheless believe that CIAD {\it should} have done so. 
They state that adding $f_e$ to the equation of vertical motion "follows logically" from \citet{g12}'s statement that
"the power of latent heat release which is $\xi$ times larger than the power of dynamic condensation-induced circulation". This "suggests"
to JMR2 that $f_e$ must be added to the list of forces in the equation of vertical motion. JMR2 characterise this self-written proposal as 
"a strong contradiction to their claim that CIAD does not modify the vertical momentum budget."

This proposition is incorrect.
A comparison of the rates of heat release and kinetic energy generation does not 
"suggest" any modification of the equations of motion.  For example, in a close parallel to \citet{g12},
\citet{emanueletal1994}, following \citet{charney64}, emphasized that in hurricanes "the latent heat energy released is two orders of magnitude greater than the amount needed 
to maintain the kinetic energy against frictional dissipation". At the same time, \citet{emanueletal1994} warned against drawing any dynamic conclusions from such a comparison
explaining that "the implication that heating per se leads to production of kinetic energy" is "an important misconception in the atmospheric science".
 We note that JMR2 have based their incorrect claim on this misconception.

\vspace{0.3cm}
{\bf 2. \boldmath $F_{vd} = f_e$.} 
A second claim from which JMR2 retreat
is that in hydrostatic equilibrium the evaporative force $f_e$ is equal to an {"internal force"} $F_{vd}$ exerted by vapor on dry air.
\citet{CIAD19} (hereafter CIAD19) pointed out that $f_e = F_{vd}$ can only be obtained by misinterpretating the equations of motion -- in particular, by believing incorrectly
that the partial pressure gradient of water vapor (dry air) acts exclusively on water vapor (dry air).

JMR2 made no attempt to defend their $f_e = F_{vd}$ (this statement was not referred to again). Instead, 
JMR2 added an identical zero $F_{dv} + F_{vd} = 0$ to the equation of vertical motion (with neither $F_{dv}$ nor $F_{vd}$ specified). This procedure lacks a specified physical  meaning but
convinced JMR2 that $f_e$ "being defined as the force that the water vapor exerts on the air is {\it covered} in the definition of $F_{vd}$" and that $f_e$ is an "internal force". Note that one can add {\it any expression}
simultaneously with the plus and minus sign and claim in a similar manner that $f_e$ is {\it covered in the definition of whatever} has been added. 
Without proper explanations and justifications such ill defined inferences serve only to confuse.

\vspace{0.3cm}
{\bf 3. $f_e$ in the equation of vertical motion} 

A new claim by JMR2 is that CIAD19 "presents a strong contradiction proposing $f_e$ as a new force that exerts power, but at the same time,
it is not present in the momentum budget". This statement is incorrect, since the fact that CIAD {\it does not add} $f_e$ to the vertical momentum budget
does not mean that {\it it is not already present there}. Indeed, it is. Given that $\rho = p/(gh)$, where $h \equiv RT/(Mg)$, $\rho$ and $p$ is air density and pressure,
$T$ is temperature, $M$ is air molar mass, we can write the equation of vertical motion as
\beq
\label{vm}
\rho \frac{Dw}{Dt} = -\frac{\pt p}{\pt z} - \rho g = -\frac{\pt p_v}{\pt z} - \frac{p_v}{h} -\frac{\pt p_d}{\pt z} - \frac{p_d}{h} \equiv f_e + f_d.
\eeq
Here $f_e \equiv -\pt p_v/\pt z -p_v/h$ the evaporative force as defined by Eq. (9) of CIAD19 and $f_d \equiv -\pt p_d/\pt z -p_d/h$.
 So again JMR2 make incorrect claims that may confuse rather than clarify the merits of CIAD.

\vspace{0.3cm}
{\bf 4. Is there a rationale in JMR's statements?} 

We have been in correspondence with JMR for an extended period (see footnotes in our detailed comments in arxiv http://arxiv.org/abs/1809.01874), so we have formulated a vision of how their
claim could have been formulated more coherently (and we had long ago addressed that claim too). Looking at Eq.~(\ref{vm}) one can ask, if
work $wf_e$ per unit time is what provides positive power to atmospheric circulation, why does $wf_d$ deriving from the dry air distribution 
not provide an equal opposite in magnitude power such that the net impact of these two forces is zero? This question does not require new terms or confusing claims.

The answer lies in noting the implicit misconception of JMR1 and JMR2 that atmospheric power relates to all the work of {\it all the forces}
in the momentum budget. The definition of atmospheric wind power $K$ (which neither JMR1 nor JMR2 discuss) is $K = -\int \mathbf{v}\cdot \nabla p d\mathcal{V}$, i.e.
the integral over the atmospheric volume of the scalar product of wind velocity $\mathbf{v}$ and pressure gradient. 
The right-hand part of the equation of vertical motion contains, besides
the vertical pressure gradient, also the gravity force; but it does not contain the horizontal pressure gradient. Whatever happens in the right-hand side of this equation,
whether the forces sum up to zero or not, does not impact the atmospheric power budget. In other words, it does not follow from the fact that $w(f_e+f_d) = 0$ that
$K = 0$. (For interested readers, in the Appendix we discuss a similar situation from the literature
when work of the buoyancy force, despite this force is like $f_e$ exactly compensated by other forces in the right-hand part of the equation of vertical motion,
nevertheless determines the available potential energy for a {\it hydrostatic} atmosphere with adiabatic motions.)

As we have discussed elsewhere, see, e.g. \citet[][their Eq. 17]{g12} and \citet[][p. 1047]{m13}, the deviation of the dry air distribution from hydrostatic equilibrium is a consequence of the fact
that the atmosphere as a whole is in hydrostatic equilibrium, $f_d + f_e = 0$, such that all power generated by the work per unit time of the evaporative
force $f_e$ is manifested in the horizontal plane. Acknowledgment, and appreciation, of this physical context would avoid the misconceptions seen in JMR1 and JMR2.

\vspace{0.3cm}
{\bf 5. What is new in JMR2: "parameterization"} 

JMR2 provide a new argument. They
do not dispute that the main dynamic equation of CIAD,
\beq
\label{ciad}u\frac{\pt p}{\pt x} = \mathscr{C}_{CIAD},
\eeq
where
\beq\label{C}\mathscr{C}_{CIAD} \equiv -wf_e \equiv w\left( \frac{\pt p^*_v}{\pt z} - \frac{p_v^*}{p} \frac{\pt p}{\pt z}\right)\eeq
is consistent with observations in different atmospheric contexts.

But now they argue that this result, Eq.~(\ref{ciad}), does not require a theory but follows
from the continuity equation and "a parameterization" of condensation rate $\mathscr{C}$
of the type
\beq
\label{CJ}
\mathscr{C}_{CIAD} = \alpha \mathscr{C}^*_z,
\eeq
where $\alpha \lesssim 1$ and $\mathscr{C}_z^*$ is given by Eq. 14 of JMR2 evaluated for the saturated vapor pressure.

This claim is incorrect. As $p_v/p \ll 1$ the continuity equation does not possess
a sufficient accuracy to determine $u\pt p/\pt x$ (see below). Thus, while the CIAD equation
is consistent with both the continuity equation\footnote{Note that while in both JMR2 and CIAD works the turbulent diffusion terms are formally neglected in the continuity equation, in
fact turbulent diffusion is implicitly present in the condition $\pt p_v/\pt x = 0$, i.e. turbulent diffusion is what ensures constant relative humidity at the surface on an isothermal plane.} and the above approximate relationship,
it cannot be derived from them.

Indeed, with $\alpha \approx 1$ the continuity Equation (17) of JMR2
\beq\label{cJ}
u\frac{\pt p}{\pt x} = \frac{p_d}{p_v^*}(1-\alpha) \mathscr{C}_z^*\eeq
contains a product of a large factor $p_d/p_v \sim 10^2$ and an unknown small factor $1 -\alpha \ll 1$.
Thus, assuming that a certain (a priori unknown) $\alpha \approx 1$ matches the observations, a mere 10\% reduction
of $\alpha$, while still obeying $\alpha \approx 1$, would lead to an order of magnitude overestimate
of $u \pt p/\pt x$. Conversely, any $p_d/p < \alpha < 1$ will produce unrealistically low (down to zero) values of $u \pt p/\pt x$.
Thus, the CIAD equation (\ref{ciad}) does not follow from the parameterization (\ref{CJ}) contrary to the statement of JMR2.

\vspace{0.3cm}
{\bf 6. Conclusions} 

Despite their misunderstandings and criticisms of CIAD, JMR2 don't dispute that
 CIAD's "main dynamic equation" is applicable to a wide range of atmospheric phenomena and "describes wind and pressure profiles
as consistent with phenomena like hurricanes and tornadoes, the circulation in the Amazon rain forest, or
even an estimation of the global circulation power that appears to be consistent with observations." Indeed, JMR2 now consider this equation "interesting" and they have even attempted
to propose their own (incorrect) explanation for why it matches the observations. 
(We note that if JMR now wanted to argue that the main CIAD equation is invalid, they would have to invalidate their own justification for it.)
Before CIAD no universal constraint on atmospheric power was known, this alone makes CIAD worthy of constructive attention  whatever views one may hold on its other merits.
We are happy with this outcome and foresee many additional constructive developments in the future. We thank JMR1 and JMR2 for their efforts and for having drawn extra attention to our work
and allowing us an opportunity to address misconceptions and clarify interpretations.

\vspace{0.6cm}
{\bf 7. Appendix "Work of buoyancy force in a hydrostatic atmosphere"} 

\citet{lorenz55,lorenz78} proposed his formulations for 
available potential energy to drive winds that involve the differences in temperature and density between local air with
pressure $p$ and density $\rho$ and a hydrostatic reference environment with pressure $\overline{p} = \overline{p}(z)$
and density $\overline{\rho} = \overline{\rho}(z)$ with
\beq\label{meanhe}
\frac{\pt \overline{p}}{\pt z} + \overline{\rho}g = 0.
\eeq
Adding Eq.~(\ref{meanhe}) to Eq.~(\ref{vm}) and 
using the definition of the buoyancy force $f_b \equiv (\overline{\rho} - \rho) g$, we find
that the equation of vertical motion (\ref{vm}) becomes
\beq\label{vm2}
\rho \frac{Dw}{Dt} = f_b  + \frac{\pt \overline{p}}{\pt z}-\frac{\pt p}{\pt z}.
\eeq
Lorenz' formulations were made for a hydrostatic atmosphere (dry or moist). Later it was shown
that Lorenz' moist available energy can be reformulated in terms
of convective available potential energy, which is equal to the work of the buoyancy force on the rising air parcels \citep[e.g.,][pp.~179-185]{emanuel94}.

However, precisely as $f_e$ (or indeed any other vertical force)
is, by definition of hydrostatic equilibrium, compensated by the sum of all the other vertical forces, $f_e + f_d = 0$ in Eq.~(\ref{vm}),
the buoyancy force $f_b$ in a hydrostatic atmosphere is also compensated by the other forces, i.e. by the last two terms in Eq.~(\ref{vm2}).
Nevertheless, despite this fact, the potential energy available for atmospheric motions is not zero. Under the approximations Lorenz made,
it is equal to the work of the (compensated) buoyancy force.
This is because in a hydrostatic atmosphere all processes in the vertical are compensated, but related in their magnitude to the 
wind power generated in the horizontal plane. For the same reason, \citet{em86} was able to formulate a description of buoyantly neutral hydrostatic hurricanes
that was closely related to, and produced similar quantitative results as, the previous descriptions based on consideration of the storm's positive buoyancy.
\large
%\bibliography{met-refs}

\begin{thebibliography}{11}
\providecommand{\natexlab}[1]{#1}
\providecommand{\url}[1]{\texttt{#1}}
\providecommand{\urlprefix}{URL }
\expandafter\ifx\csname urlstyle\endcsname\relax
  \providecommand{\doi}[1]{doi:\discretionary{}{}{}#1}\else
  \providecommand{\doi}{doi:\discretionary{}{}{}\begingroup
  \urlstyle{rm}\Url}\fi
\providecommand{\eprint}[2][]{\url{#2}}


\bibitem[{Charney~and Eliassen(1964)}]{charney64}
Charney, J.~G., and A.~Eliassen, 1964. 
On the growth of the hurricane depression. \textit{J. Atmos. Sci.}, \textbf{21},
  68--75, \doi{10.1175/1520-0469(1964)021$<$0068:OTGOTH$>$2.0.CO;2}.

\bibitem[{Emanuel(1986)}]{em86}
Emanuel, K.~A., 1986: An air-sea interaction theory for tropical cyclones.
  {Part I: Steady-state} maintenance. \textit{J. Atmos. Sci.}, \textbf{43},
  585--604, \doi{10.1175/1520-0469(1986)043$<$0585:AASITF$>$2.0.CO;2}.

\bibitem[{Emanuel(1994)}]{emanuel94}
Emanuel, K.~A., 1994: \textit{Atmospheric convection}. Oxford University, 592
  pp.

\bibitem[{Emanuel et~al.(1994)Emanuel, Neelin, and
  Bretherton}]{emanueletal1994}
Emanuel, K.~A., J.~D. Neelin, and C.~S. Bretherton, 1994: On large-scale
  circulations in convecting atmospheres. \textit{Q. J. R. Meteorol. Soc.},
  \textbf{120}, 1111--1143.

\bibitem[{Gorshkov et~al.(2012)Gorshkov, Makarieva, and Nefiodov}]{g12}
Gorshkov, V.~G., A.~M. Makarieva, and A.~V. Nefiodov, 2012: Condensation of
  water vapor in the gravitational field. \textit{J. Exp. Theor. Phys.},
  \textbf{115}, 723--728, \doi{10.1134/S106377611209004X}.

\bibitem[{Jaramillo et~al.(2018)Jaramillo, Mesa, and Raymond}]{JMR1}
Jaramillo, A., O.~J. Mesa, and D.~J. Raymond, 2018: Is condensation-induced
  atmospheric dynamics a new theory of the origin of the winds? \textit{J.
  Atmos. Sci.}, \textbf{75}, 3305--3312, \doi{10.1175/JAS-D-17-0293.1}.

\bibitem[{Jaramillo et~al.(2019)Jaramillo, Mesa, and Raymond}]{JMR2}
Jaramillo, A., O.~J. Mesa, and D.~J. Raymond, 2019: Reply to "comments on '{Is}
  condensation-induced atmospheric dynamics a new theory of the origin of the
  winds?'’’. \textit{J. Atmos. Sci.}, \textbf{76}, 2187--2191,
  \doi{10.1175/JAS-D-19-0025.1}.

\bibitem[{Lorenz(1955)}]{lorenz55}
Lorenz, E.~N., 1955: Available potential energy and the maintenance of the
  general circulation. \textit{Tellus}, \textbf{7}, 157--167,
  \doi{10.3402/tellusa.v7i2.8796}.

\bibitem[{Lorenz(1978)}]{lorenz78}
Lorenz, E.~N., 1978: Available energy and the maintenance of a moist
  circulation. \textit{Tellus}, \textbf{30}, 15--31,
  \doi{10.3402/tellusa.v30i1.10308}.

\bibitem[{Makarieva et~al.(2019)Makarieva, Gorshkov, Nobre, Nefiodov, Sheil,
  Nobre, and Li}]{CIAD19}
Makarieva, A.~M., V.~G. Gorshkov, A.~D. Nobre, A.~V. Nefiodov, D.~Sheil,
  P.~Nobre, and B.-L. Li, 2019: Comments on ‘‘{Is} condensation-induced
  atmospheric dynamics a new theory of the origin of the winds?’’.
  \textit{J. Atmos. Sci.}, \textbf{76}, 2181--2185,
  \doi{10.1175/JAS-D-18-0358.1}.

\bibitem[{Makarieva et~al.(2013)Makarieva, Gorshkov, Sheil, Nobre, and
  Li}]{m13}
Makarieva, A.~M., V.~G. Gorshkov, D.~Sheil, A.~D. Nobre, and B.-L. Li, 2013:
  Where do winds come from? {A} new theory on how water vapor condensation
  influences atmospheric pressure and dynamics. \textit{Atmos. Chem. Phys.},
  \textbf{13}, 1039--1056, \doi{10.5194/acp-13-1039-2013}.

\end{thebibliography}

\end{document}